\documentclass{icrc2009}

\usepackage{graphicx}   
\usepackage[caption=false]{caption}    
\usepackage[font=footnotesize]{subfig} 
\usepackage{fixltx2e}
\usepackage{url}

\newcommand{\shorttitle}[1]%
{\markboth{Proceedings of the 31\MakeLowercase{$^{st}$} ICRC, {\L}\'{o}d\'{z} 2009}{#1} }
\newcommand{\etal}{\MakeLowercase{\textit{et al. }}} 

\begin{document}
\title{Localized Galactic sources and their contribution beyond the second knee}

\author{\IEEEauthorblockN{Cinzia De Donato\IEEEauthorrefmark{1} and 
			  Gustavo Medina-Tanco\IEEEauthorrefmark{1}
}
                            \\
\IEEEauthorblockA{\IEEEauthorrefmark{1}Dep. Altas Energias, Inst. de Ciencias Nucleares, Universidad Nacional Autonoma de M\'exico, M\'exico DF, CP 04510.}
}

\shorttitle{C. De Donato \etal High Energy CR Galactic Sources}
\maketitle

\begin{abstract}
The energy range encompassing the ankle of the cosmic ray energy spectrum probably
marks the exhaustion of the accelerating sources in our Galaxy, as well as the end
of the Galactic confinement. Furthermore, this is the region where the extragalactic
flux penetrates the interstellar medium and starts, progressively, to be dominant.
Although at lower energies it is likely that an ``average" population of supernova
remnants can be defined to account for most of the cosmic ray flux, this assumption
is increasingly difficult to maintain as higher energies are considered. One
possibility is that supernovas are still a main contributor along the first branch of
the ankle region, but that the acceleration is now coming from well localized
regions with a characteristic interstellar medium, or a sub-population of supernovas
exploding in a peculiar circumstellar environment. These possibilities are analyzed in
the present work using a two-dimensional diffusion model for cosmic ray
propagation. Special emphasis is given to the inner 200 pc of our Galaxy and to the
 spiral arm structure in relation with the Sun position inside the disk. 
\end{abstract}

\section{Introduction}
In our previous work \cite{DeDonato2007, DeDonato2008}, we have analyzed the matching conditions of the Galactic and
extragalactic components of cosmic rays (CR) along the second knee and
the ankle. From this analysis, it seems clear that an acceptable matching of the Galactic and
extragalactic fluxes can only be achieved if the Galaxy has
additional accelerators, besides the fiducial Supernova Remnants (SNRs) assumed there, 
operating in the interstellar medium. Despite the fact that a different acceleration mechanism can be invoked to account 
 for the highest energy side of the Galactic spectrum,  acceleration by SNRs may still play a dominant role. Actually, 
a possible Galactic contribution, dominating at the highest energies, could be represented by 
 compact and highly magnetized SNRs, like those occurring in the central, high density regions of
the Galactic bulge, inside the dense cores of molecular clouds or those expanding
into the circumstellar winds of their progenitors. In the same way, the highest energy end of the Galactic spectrum 
 could be the result of a non-homogeneous SNR population drawn from a spectrum of progenitor masses and evolving in different
environments corresponding to the various gas phases that fill the interstellar
medium \cite{Stanev1993}.\\
These possibilities are explored in this paper calculating with a  diffusion Galactic model the contribution of localized sources to  the Galactic cosmic ray (GCR) spectrum,
 SNRs localized in the Galactic center (GC)  and in the Galactic Ring presenting the highest CO emissivity.
These two contributions, along with the Galactic spectrum resulting from fiducial SNRs (called \emph{standard} in the following), 
 are combined with a mixed extragalactic (EG) spectrum and normalized to match HiRes \cite{HiRes2001,HiRes2005} and Auger  
 \cite{AugerSD2007,AugerHybrid2007} experimental data.
The EG model considered is the mixed composition model by Allard \etal \cite{Allard2007} in the case of the uniform source distribution model.

\section{SNR evolution}
The maximum acceleration energy by SNRs is related to the  shock radius  at the end of the adiabatic phase.
In this section we estimate the radius and age of a blastwave at the end of its adiabatic phase for two different characteristic cases,
  a uniform ambient density and a power-law density gradient $\rho\propto r^{-2}$, the latter corresponding to explosions in a pre-existing supersonic wind.\\
The dynamic of the blastwave in  stationary and homogeneous media, is described by the self-similar  Sedov-Taylor (ST) solution, valid in an 
 ambient with negligible pressure $P_0\simeq0$ \cite{Ostriker1988}.
In media with power-law density variations $\rho_0(r)\propto r^{-k_\rho}$ 
(for $k_\rho\leq k_{\rho crit}$\begin{footnote}{$k_{\rho crit}=\frac{7-\gamma}{\gamma+1}$, where $\gamma$ is the specific heats 
 ratio at the shock.}\end{footnote}), 
 the evolution of a blastwave of initial energy $E_0$ in the adiabatic phase is given by 
\setlength{\arraycolsep}{0.0em}
\begin{eqnarray}
\label{RsAdiab}
R_s&=&R_s(1)\left[ \frac{\xi E_0 t^2}{\bar{\rho}(1) R_s(1)^5}\right]^{\eta/2}=\\
&=&\left[ \frac{1.52 \times 10^{-3} (1-k_\rho/3)\xi E_{51}}{n_0(1)} \left(\frac{t}{yr}\right)^2 \right]^{\eta/2}~pc,\nonumber
\end{eqnarray}
\setlength{\arraycolsep}{5pt}
\noindent where $R_s$ is the radius of the blastwave (normalized to some fiducial value $R_s(1)$), $E_{51}=\frac{E_0}{10^{51} erg}$ and $n_0(1)$ is the 
 ambient hydrogen number density in $cm^{-3}$ at $R_s(1)$.
The quantity $\xi$  in eq. \ref{RsAdiab} is given by
\begin{equation}
\label{xi}
\xi=\frac{3}{4\pi\eta^2\sigma},
\end{equation}
where  $\eta= \frac{2}{5-k_\rho}$ and $\sigma$ is the ratio between the total energy of the gas inside the blastwave and its kinetic energy.
For a specific heat ratio $\gamma=\frac{5}{3}$, $\xi$ assumes the value
\begin{equation}
\xi=\frac{3(5-k_\rho) (10-3k_\rho)}{8\pi(3-k_\rho)}.
\end{equation}

\noindent The age $t$ correspondent to the shock radius $R_s$ is 
\begin{equation}
\label{tAdiab}
t=25.6\left( \frac{n_0(1)}{(1-k_\rho/3)\xi E_{51}}\right)^{1/2}\left(\frac{R_s}{pc}\right)^{2/\eta}~yr.
\end{equation}

\noindent The adiabatic phase ends when radiative cooling  dominates the blastwave evolution. 
The transition point can be estimated as the radius $R_c$ (time $t_c$) at which the half of the initial energy has been radiated away:
\setlength{\arraycolsep}{0.0em}
\begin{eqnarray}
\label{Rc}
R_c&=&\left[\frac{ (7-3 k_\rho)(3-k_\rho)^2 \eta^2 \xi E_{51}^2 }{n_0(1)^3 I_{-1/2} } \right]^{1/(7-k_\rho)}\\\nonumber
 &\times&\left[4.96 \times 10^{8}\right]^{1/(7-k_\rho)}~pc,
\end{eqnarray}
\setlength{\arraycolsep}{5pt}
\noindent where $I_{-1/2}$ is the radiative form factor for line-cooling. 
The corresponding age of the blastwave $t_c$ can be estimated using eq. \ref{tAdiab}.

\noindent From this equation the radius and age of blastwaves at the end of the adiabatic phase can be calculated for our two cases (for $\gamma=5/3$):
\begin{itemize}
\item uniform density, $k_\rho=0$
 \setlength{\arraycolsep}{0.0em}
\begin{eqnarray}
\label{k0}
R_c&=&24.6 \left[ E_{51}^2 n_0(1)^{-3}\right]^{1/7}~pc,\\
t_c&=&5.4 \times 10^4 \left[E_{51}^{3/2} n_0(1)^{-4}\right]^{1/7}~yr;
\end{eqnarray}
 \setlength{\arraycolsep}{5pt}
\item power-law density, $k_\rho=2$
 \setlength{\arraycolsep}{0.0em}
\begin{eqnarray}
\label{k2}
R_c&=&7.9 \times 10^7 \left[ E_{51}^2 n_0(1)^{-3}\right]~pc,\\
t_c&=&2.6 \times 10^{13} \left[E_{51}^{5/2} n_0(1)^{-4}\right]~yr.
\end{eqnarray}
 \setlength{\arraycolsep}{5pt}
\end{itemize}

\noindent A uniform density ISM is a good approximation for SNRs evolving inside the GC region, as well as for the main SNR component (``standard") expanding inside the general ISM elsewhere in the Galactic disk. In fact the ISM inside the inner 200 pc of the Galaxy has a density of $\sim 10^{4}$ cm$^{3}$, which is very similar to that of the the cores of molecular clouds, but extended over a large region where several compact remnants can be easily observed in radio at different stages of evolution \cite{GMT2007}. Furthermore, this region is permeated by a magnetic field almost three orders of magnitude larger than in the rest of the Galactic disk. A power density variation of the form $r^{-2}$ (i.e., $k_\rho=2$), on the other hand, corresponds to a SN blast wave propagating through the circumstellar wind of its progenitor. It is unlikely that these circumstellar regions could preserve such a well structured density profile inside the high density, high turbulence environment of the Galactic center. Nevertheless, a  $k_\rho=2$ profile could described well the ambient medium of SNR expanding inside the still dense, but rather isolated regions, associated with dispersed molecular clouds inside the Ring.\\

\noindent The maximum acceleration energy for a particle of charge Z  is proportional to  $ZeBR$, where $B$ and $R$ are  the magnetic field  intensity and the size of the acceleration region. If we consider  a standard ISM and the GC, typical values for  gas density and magnetic field are $n_0 \sim 1$ cm$^{-3}$ and $B\sim \mu G$, for the first and $n_0 \sim 10^4$ cm$^{-3}$ and $B\sim mG$ for the second. The maximum scale $R$ associated with particle acceleration can be estimated as the SNR radius at the transition point between the adiabatic and the radiative phase, given by eq. \ref{k0} or \ref{k2}, depending on the value of $k_{\rho}$.  

\noindent The ratio between the maximum energy achievable by protons by acceleration in SNRs in  the two cases is:
\begin{equation}
\frac{E_{max}^{GC}}{E_{max}^{st}}\sim \frac{B_0^{GC}}{B_0^{st}} \times  \frac{R_c^{GC}}{R_c^{st}}.
\end{equation}
In the case of uniform gas density, using eq. \ref{k0}, the ratio becomes
\begin{equation}
\label{GC/st}
\frac{E_{max}^{GC}}{E_{max}^{st}}\sim 20.
\end{equation}

\section{Diffusion Galactic model\label{sec:GalacticModel}}

We used the numerical diffusive propagation code
GALPROP \cite{Strong1998,Strong2001} to reproduce the galactic
spectrum from SuperNova Remnants (SNRs). The diffusive model is
axisymmetric. The propagation region is, in cylindrical coordinates,
bounded by $R=R_h=30~kpc$ and $z=z_h=4~kpc$, beyond which free
escape is assumed, where $R$ is the Galactocentric distance and $z$ is the altidude from the Galactic plane.\\
The propagation equation is:

\vskip -0.3cm
\begin{eqnarray}\label{diffeq}
\frac{\partial\psi}{\partial t} & = & q(\vec{r},p)+ \vec{\nabla}
\cdot (D_{xx}\vec{\nabla} \psi)+ \nonumber \\
& & -\frac{\partial}{\partial
p}(\dot{p}\psi)-\frac{1}{\tau_f}\psi-\frac{1}{\tau_r}\psi
\end{eqnarray}

\noindent where $\psi(\vec{r},p,t)$ is the density per unit of total particle
momentum, $q(\vec{r},p)$ is the source term, $D_{xx}$ is the spatial
diffusion coefficient, $\dot{p}= dp/dt$ is the momentum loss rate
and $\tau_f$ and $\tau_r$ are the time scale of fragmentation and
the time scale of radioactive decay respectively. The diffusion
coefficient is taken as $\beta D_0(\rho / \rho_D)^{\delta}$ , where
$\rho$ is the particle rigidity, $D_0$ is the diffusion coefficient at a reference rigidity $\rho_D$ and $\delta=0.6$.\\
Stable nuclei with $Z<26$ are injected, at the source with energy independent isotopic
abundances derived from low energy CR measurements \cite{Strong2001}.\\
Detailed and realistic interstellar molecular (H$_2$), atomic (H)
and ionized (HI) hydrogen distributions are used \cite{Moskalenko2002}.

\noindent The distribution of cosmic rays sources used for \emph{standard} SNRs  is 
that of Galactic SNRs deduced from  EGRET gamma-ray data \cite{Strong1998}, while for SNRs operating in the 
GC and in the Ring we used a uniform distribution of sources limited to the regions $R\leq0.2$ kpc, $|z|\leq0.2$ kpc 
 and $4.5\leq R\leq5.5$ kpc, $|z|\leq0.1$ kpc, respectively.
 The injection spectrum is a a power law function
in rigidity with a break at rigidity $\rho_0$, beyond which it falls
exponentially with a rigidity scale $\rho_c$:

\begin{displaymath}
I(\rho)= \left\{ \begin{array}{ll}
\label{InjSp}
 \left(\frac{\rho}{\rho_0}\right)^{-\alpha} & \rho\leq\rho_0\\
 \left(\frac{\rho}{\rho_0}\right)^{-\alpha} \exp\left[-\frac{\rho- \rho_0}{\rho_c}\right] & \rho>\rho_0 \nonumber
\end{array}\right.
\end{displaymath}
\noindent where $\alpha=2.05$, while $\rho_0$ and $\rho_c$ depend on the SNRs considered:
\begin{itemize}
\item \emph{standard} SNRs: $\rho_0^{st}=3.6$ PV, $\rho_c^{st}=2.5$ PV;
\item Ring SNRs: $\rho_0^R=17.5$ PV, $\rho_c^R=12.15$ PV;
\item GC SNRs: $\rho_0^{GC}=80$ PV, $\rho_c^{GC}=55.6$ PV.
\end{itemize}

The rigidity scale and rigidity cut-off for \emph{standard} SNRs were determined empirically so that the total GCR spectrum fits the position of the first knee measured by KASCADE ($E_{knee}\sim 5$ PeV) \cite{Kascade2005} and the shape of the observed spectrum beyond it up to the highest possible energies.
 The estimate of the rigidity cut-off for the SNRs in the GC and in the Ring were also determined empirically but in order to fit the measured total spectrum, once combined with the standard SNRs spectrum and with the EG spectrum. 
 This estimate was performed taking into account the estimate of $E_{max}^{GC}/E_{max}^{st}$ given by eq. \ref{GC/st}, 
 which suggests a rigidity cut-off for the GC component   $\rho_{0}^{GC}\sim20 \times{\rho_{0}^{st}}$.\\
The physical conditions required for the Ring region in order to provide the rigidity cut-off  $\rho_0^R=17.5$ PV can be estimated from eq. \ref{k2}.\\
The ratio between the maximum energy achievable by protons  in the case of SNRs exploding in a pre-existing wind in  
 the Ring region  and \emph{standard} SNRs  is:
\begin{equation}
\frac{E_{max}^{Ring}}{E_{max}^{st}}\sim \frac{B_0^{Ring}}{B_0^{st}} \times  \frac{R_c^{Ring}}{R_c^{st}}.
\end{equation}
For a magnetic field density $\sim 10 \mu G$ characteristic of the Ring region and the typical values of the Galactic  ISM,
 using eqs. \ref{k0}, \ref{k2} and our empirical ratio  $\rho_0^R/\rho_0^{st}\sim5$, the required gas density in the Ring region 
 is of order of $200$ cm$^{-3}$. This density value is reasonable for the Ring region, characterized by heterogenous medium of regular ISM interspersed with molecular cloud cores of density as high as $10^4$-$10^5$ cm$^{-3}$.

\section{Diffusive Galactic spectrum}

\begin{figure*}[th]
  \centering
  \includegraphics[width=5in]{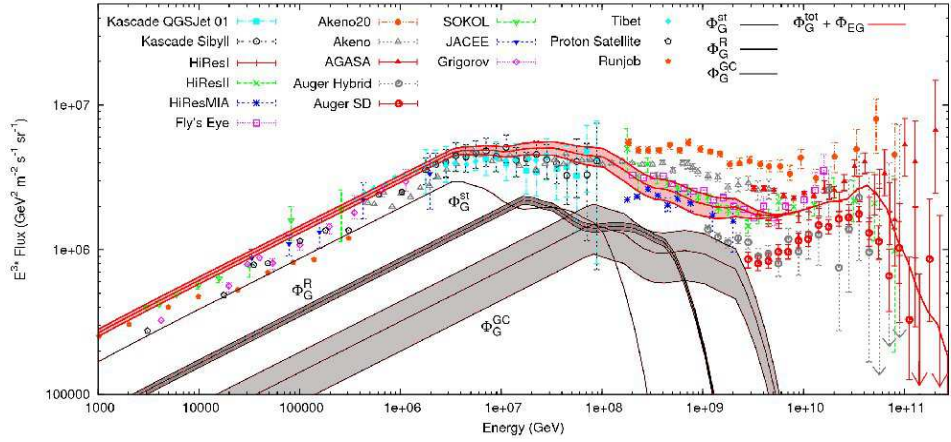}
  \caption{Diffusive total Galactic spectrum ($\Phi_{G}^{tot}$) combined with the mixed EG spectrum ($\Phi_{EG}$) and normalized to 
 KASCADE \cite{Kascade2005} and HiRes data \cite{HiRes2001,HiRes2005}.  
The different contributes  from \emph{standard} SNRs ($\Phi_{G}^{st}$), SNRs in the GC ($\Phi_{G}^{GC}$)  and in the Ring ($\Phi_{G}^{R}$) are shown. 
 The curves limiting the hatched areas correspond to a variation $\pm \sqrt{N}$ of the number of contributing sources in the GC and in the Ring.
HiRes data and other several experimental data results are shown. 
}
  \label{Fig01}
 \end{figure*}

 The calculated diffusive Galactic spectra from the three sources  are combined with the EG spectrum 
 and normalized in order to match KASCADE data  at $E\sim 3\times 10^6$ GeV \cite{Kascade2005}
 and HiRes data \cite{HiRes2001,HiRes2005} at higher energy.
 The resulting CR spectrum is shown in Fig. \ref{Fig01} superimposed to several experimental data results.
 
\noindent We calculated the number of sources contributing to the different Galactic flux components. 
 Integrating the source distribution functions and assuming that the total energy pumped by a SNR into the CR component 
 is a constant fraction of its total kinetic energy  independently on its environment,
 we found that  $\sim6500$ \emph{standard} SNRs are required in order to account for the main component of the CR Galactic flux up to the second knee energy,
   while the  fluxes coming from the GC and from the Ring requires $\sim7$ and $\sim260$ sources, respectively. 
The number of required SNRs in the GC region is sufficiently small to be readily supplied by the observed population inside the inner $200$ pc of our Galaxy.
In fact, such a small number could pose a potential problem, since random variations in the number of SNRs could distort the shape of the 
 spectrum along the ankle as a function of time.\\
 In order to check the extent of their possible effect, we have perturbated the number $N$ of SNRs in the GC population by 
$\pm \sqrt{N}$ (and analogously for the number of sources in the Ring). The results are shown in Fig. \ref{Fig01} by the curves limiting the hatched areas, where it can be seen that the ankle remain smooth but, under extreme conditions, its location could change in energy by up to half a decade under stochastic fluctuations of the population of high energy SNRs.
In any case, since the average time between SN in the GC is much smaller than the $10^4$ yr of diffusion time up to the solar circle, 
 the time dependence of the ankle should be further suppressed.

\begin{figure*}[!t]
  \centering
  \includegraphics[width=5in]{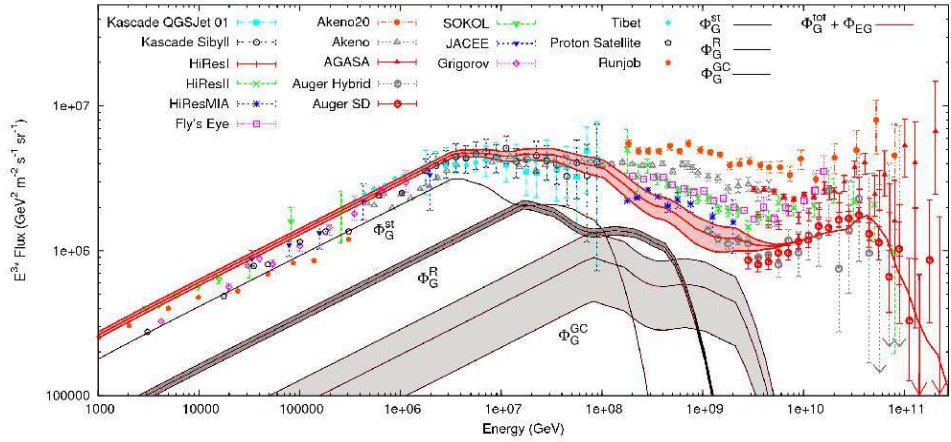}
  \caption{Diffusive total Galactic spectrum ($\Phi_{G}^{tot}$) combined with the mixed EG spectrum ($\Phi_{EG}$) and normalized to  
 KASCADE \cite{Kascade2005} and Auger \cite{AugerHybrid2007, AugerSD2007} data.  
The different contributes  from \emph{standard} SNRs ($\Phi_{G}^{st}$), SNRs in the GC ($\Phi_{G}^{GC}$)  and in the Ring ($\Phi_{G}^{R}$) are shown. 
  The curves limiting the hatched areas correspond to a variation $\pm \sqrt{N}$ of the number of contributing sources in the GC and in the Ring.
Auger data and other several experimental data results are shown. 
}
  \label{Fig02}
 \end{figure*}

\noindent The same procedure of normalization of the Galactic components has been applied in order to match the total spectrum  with  Auger 
 data \cite{AugerHybrid2007, AugerSD2007} at high energy and to KASCADE data at $E\sim 3\times 10^6$ GeV \cite{Kascade2005}.\\
The resulting total spectrum and each Galactic contribution are shown in Fig. \ref{Fig02}. 
In this case, the CR flux from sources in the GC is considerably lower with respect to  the HiRes case, 
 while the Ring and \emph{standard} components are of the same order of magnitude.
Integrating the source distribution functions, the number of sources contributing to the GCR flux is $\sim6900$ for  \emph{standard} SNRs
 and $\sim230$ and $\sim4$  for the Ring and GC, respectively.
As in the previous case, the result for a perturbation of the number of sources in the GC region 
and in the Ring  are indicated by the curves limiting the hatched area \ref{Fig02}. 
\noindent  Even if the number of sources in the GC are smaller with respect to the HiRes case, the global behavior is consistent with that observed in the HiRes case.

\section{Conclusion}

A previous analysis \cite{DeDonato2007, DeDonato2008} suggests 
that additional acceleration mechanisms are required besides acceleration from fiducial (\emph{standard}) SNRs in order to account for GCR flux up to the highest energy. 
In a scenario where accelerators different from SNRs are disregarded, the highest energy part of the galactic spectrum could be the result of SNRs evolving in peculiar environments. We explore here the possible contribution of SNRs inside the GC and in the dense Galactic Ring localized at $4.4\leq R\leq 5.5$ kpc. 
We have estimated empirically the maximum energy, rigidity cut-off and rigidity scale required of these additional components in order to fit the total measured spectrum. Using the Galactic diffusion model GALPROP, we have verified that these parameters are compatible with the evolutionary properties and expected CR luminosities of SNR in these two regions.  
Therefore, the present analysis suggests that acceleration by SNRs has the potential to account for the whole Galactic cosmic ray flux if three general populations are considered: (i) a main population of standard SNRs which evolve in media with $n_{0} \sim 1$ cm$^{-3}$ and $B\sim 1 \mu$G, 
(ii) $\sim 2 \times 10^{2}$ SNRs associated with ISM of $n_{0} \sim 10^{2}$ cm$^{-3}$ and $B\sim 10 \mu$G inside the inner Ring and (iii) $< 10$ SNRs immersed in the GC ISM, where $n_{0} \sim 10^{4}$ cm$^{-3}$ and $B\sim 1 $mG.

\section*{Acknowledgements}
This work is partially supported by the Mexican agencies CONACyT and UNAM's CIC and PAPIIT.

\end{document}